\documentclass{jfm}

\usepackage{graphicx}
\usepackage{natbib}

\usepackage{amsmath}
\usepackage{subcaption}
\usepackage{bm}

\usepackage{color}

\ifCUPmtlplainloaded \else
  \checkfont{eurm10}
  \iffontfound
    \IfFileExists{upmath.sty}
      {\typeout{^^JFound AMS Euler Roman fonts on the system,
                   using the 'upmath' package.^^J}%
       \usepackage{upmath}}
      {\typeout{^^JFound AMS Euler Roman fonts on the system, but you
                   dont seem to have the}%
       \typeout{'upmath' package installed. JFM.cls can take advantage
                 of these fonts,^^Jif you use 'upmath' package.^^J}%
      }
  \else
  \fi
\fi

\ifCUPmtlplainloaded \else
  \checkfont{msam10}
  \iffontfound
    \IfFileExists{amssymb.sty}
      {\typeout{^^JFound AMS Symbol fonts on the system, using the
                'amssymb' package.^^J}%
       \usepackage{amssymb}%
       \let\le=\leqslant  
       \let\ge=\geqslant  
      }{}
  \fi
\fi

\ifCUPmtlplainloaded \else
  \IfFileExists{amsbsy.sty}
    {\typeout{^^JFound the 'amsbsy' package on the system, using it.^^J}%
     \usepackage{amsbsy}}
    {\providecommand\boldsymbol[1]{\mbox{\boldmath $##1$}}}
\fi

\providecommand\bnabla{\boldsymbol{\nabla}}

\newcommand{\smt}[1]{\textcolor{black}{#1}}
\newcommand{\jbm}[1]{\textcolor{black}{#1}}
\newcommand{\smtr}[1]{\textcolor{black}{#1}}
\newcommand{\jbmr}[1]{\textcolor{black}{#1}}

\newsavebox{\astrutbox}
\sbox{\astrutbox}{\rule[-5pt]{0pt}{20pt}}

\usepackage{color}
\definecolor{light-gray}{gray}{0.5}
\definecolor{blue}{rgb}{0.0,0.0,1.0}
\definecolor{green}{rgb}{0.0,0.5,0.0}
\definecolor{red}{rgb}{1.0,0.0,0.0}
\definecolor{cyan}{rgb}{0.0,0.75,0.75}
\definecolor{magenta}{rgb}{0.75,0.0,0.75}
\definecolor{yellow}{rgb}{0.75,0.75,0.0}

\newcommand{\grad}{\nabla}
\newcommand{\pd}{\partial}

\title[GQL for 3D rotating Couette Flow]{Three-Dimensional Rotating Couette Flow Via The Generalised Quasilinear Approximation}

\author[S. M Tobias and J.B. Marston]%
{S.M. Tobias$^1$%
  \thanks{Email address for correspondence: smt@maths.leeds.ac.uk},\ns
and J.B. Marston$^2$}

\affiliation{$^1$ Department of Applied Mathematics, University of Leeds, Leeds LS2 9JT, UK\\
$^2$ Department of Physics, Brown University, Providence RI 02912-1843. USA}

\pubyear{2010}
\volume{650}
\pagerange{119--126}
\date{?; revised ?; accepted ?. - To be entered by editorial office}
\begin{document}

\maketitle

%abstract of no more than 250 words, which provides a summary of the main aims and results. 
\begin{abstract}
We examine the effectiveness of the Generalised Quasilinear (GQL) Approximation introduced by  \citet{mct16}. \smtr{This approximation splits the  variables into large and small scales in directions where there is a translational symmetry and removes  nonlinear
interactions involving only small scales.} We utilise as a paradigm problem three-dimensional, turbulent, rotating Couette flow. We compare the results obtained from Direct Numerical Solution of the equations with those from Quasilinear (QL) and GQL calculations. In this
three-dimensional setting, there is a choice of cut-off wavenumber for the GQL approximation both in the streamwise and in the spanwise directions. \smtr{We demonstrate that the GQL approximation significantly improves the accuracy of mean flows, spectra and two-point correlation functions over models that are quasilinear in any of the translationally invariant directions, even if only a few streamwise and spanwise modes are included.} We argue that this provides significant support for a programme of Direct Statistical Simulation utilising the GQL approximation.
\end{abstract}


\section{\label{sec:intro}Introduction}

The instability of shear flows and their subsequent turbulent dynamics is a fundamental problem of fluid mechanics (see e.g. \citet{sh00,dr04}). 
The roles of stratification and rotation in aiding or suppressing instabilities and modifying the turbulence have been extensively studied, owing to their importance in geophysics and astrophysics. These investigations have utilised theoretical, computational and experimental techniques to shed light on these problems and it is now fair to say that the Taylor-Couette system, the rotating Poiseuille system  and the rotating Couette system constitute important paradigm problems for understanding the interaction of rotation both with instabilities and with fully developed turbulence.

Since the pioneering work of \citet{taylor23}  much experimental and computational resource has been allocated to Taylor-Couette (TC) flow (see e.g. \citet{koschmieder93}). It is therefore interesting that comparatively little attention has been focussed on Rotating Plane Couette (RPC) flow, which can be shown to be equivalent to TC-flow in the narrow gap limit \citep{fe2000}, despite it being the simplest model problem in which rotation interacts with shear instabilities. This is no doubt due to the difficulty in designing and constructing experiments for the RPC-flow. \smtr{Despite these difficulties,} pioneering experiments {\it have} been performed by the KTH-group \citep{ta92,hat07,tta10}, who have demonstrated a bewildering array of dynamics (similar to that exhibited in TC-flow), with 17 different flow regimes being identified.

Much attention has focussed on the anticyclonic case, where rotation is destabilising, as this leads to interesting interactions and dynamics. In particular anticyclonic rotation drives secondary flows that compete with the turbulence in transporting momentum via Reynolds stresses. This competition has been investigated numerically \citep{ba96,ba97} --- to a much lesser extent than TC-flow. \smt{Recently \citet{se2015} have extended this pioneering numerical work to more turbulent flows and have demonstrated that, for a given value of the Reynolds number, the wall shear stress is a non-monotonic function of rotation and that the momentum flux has contributions from both two-dimensional and three-dimensional structures}. We also note here that the RPC-system has also been utilised to latch onto fully nonlinear states in non-rotating Couette flow in the pioneering work of \citet{n98}.

Here, rather than investigate the transitions that occur in RPC flow, we shall utilise this system as a paradigm problem for the interaction of  wall bounded shear flows and rotation to test the effectiveness of the Generalised Quasilinear (GQL) approximation for three-dimensional systems. Previous evaluations of the GQL approximation have been performed for two dimensional systems. First \citet{mct16} examined GQL in the context of two-dimensional driven turbulence on a spherical surface and $\beta$-plane and showed it to be more effective than the regular QL approximation in reproducing both the dynamics and statistics of these flows away from equilibrium. This remains true even if only a single extra mode is retained in the large scales. \citet{chmt16} also demonstrated the effectiveness of the GQL approximation for the case of axisymmetric (two-dimensional) helical magnetorotational instability (HMRI) at low magnetic Reynolds number. They showed that the GQL approximations performed significantly better in the quasilinear approximation in reproducing the statistics of this turbulent magnetohydrodynamic flow even for a fairly modest set of modes ($3-5$ modes) retained in the large scales.   

The testing of the effectiveness of the GQL approximation represents a vital piece in the formulation of  Direct Statistical Simulation (DSS), a programme for simulating the low-order statistics of turbulent flows directly.  Many fluid systems in geophysics, astrophysics and experiments are highly turbulent and involve the interaction of non-trivial mean flows with turbulent fluctuations. It has been argued that for such flows a statistical representation of the behaviour may in certain circumstances be preferable to one based on the calculation of detailed dynamics \citep{lorenz67}. In many situations such statistical representations are based either implicitly or explicitly on the quasilinear approximation, which can be shown to be inadequate as the system moves away from statistical equilibrium \citep[see e.g.][]{tobias2013direct}. \jbmr{By extending the quasilinear approximation, accurate application in a more extensive parameter regime is possible, providing insight into key processes such as wave -- mean-flow interactions.}  

\jbmr{Another reason for developing statistical representations is that such methods may provide a route to the construction of subgrid models for unresolved dynamics that can be systematically improved.  As the short wavelength modes in the GQL approximation interact only with the long wavelength modes, and not amongst themselves, they can be replaced exactly by a generalized second-order closure \citep{mct16} called GCE2.  A test of the accuracy of GQL is therefore also a test of the accuracy of the GCE2 form of DSS.  The extension and testing of the approximation to three dimensional flows is crucial to the program of DSS; not only are the three-dimensional flows that we discuss here extremely anisotropic in the streamwise and spanwise direction, they are also characterised by the presence of a forward turbulent cascade in contrast with the two dimensional models that have been investigated previously.}

\jbmr{It is also interesting to relate the GQL approximation to a class of  restricted nonlinear (RNL) models that have been utilised in the description of (non-rotating) plane Couette flow \citep{thomasea2014,tfig2015,bretheimea2015}. These models have been shown to reproduce the mean profiles of plane Couette flow, including a log-layer.  RNL corresponds to a particular limit of GQL that is quasilinear in one of the translationally invariant directions and fully nonlinear in the other.}

In the next Section we present the formulation of the problem and introduce the GQL approximation for three-dimensional systems. In Section \ref{sec:results} we describe the dynamics and statistics of weakly and moderately rotating Couette flow as revealed by Direct Numerical Simulation and proceed to evaluate how well the QL and GQL approximation can capture the low-order statistics of this evolution. We conclude with a discussion in Section \ref{sec:end}.

\section{\label{sec:formulation}Formulation}
\subsection{Model, Equations and Numerical Method.}

\begin{figure}
 \centering
   \includegraphics[width=0.89\textwidth]{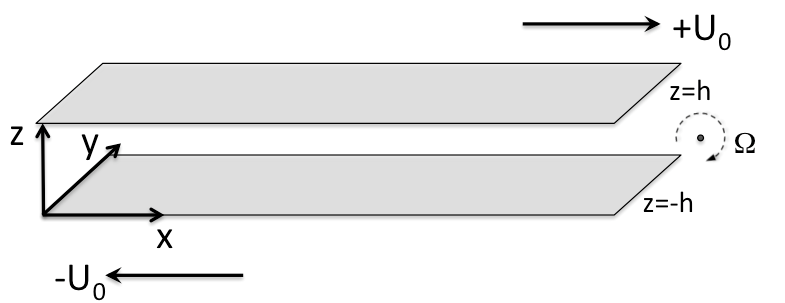}
  \caption{Rotating Couette set-up.}
  \label{fig:setup}
 \end{figure}

\begin{figure}
 \centering
   \includegraphics[width=0.89\textwidth]{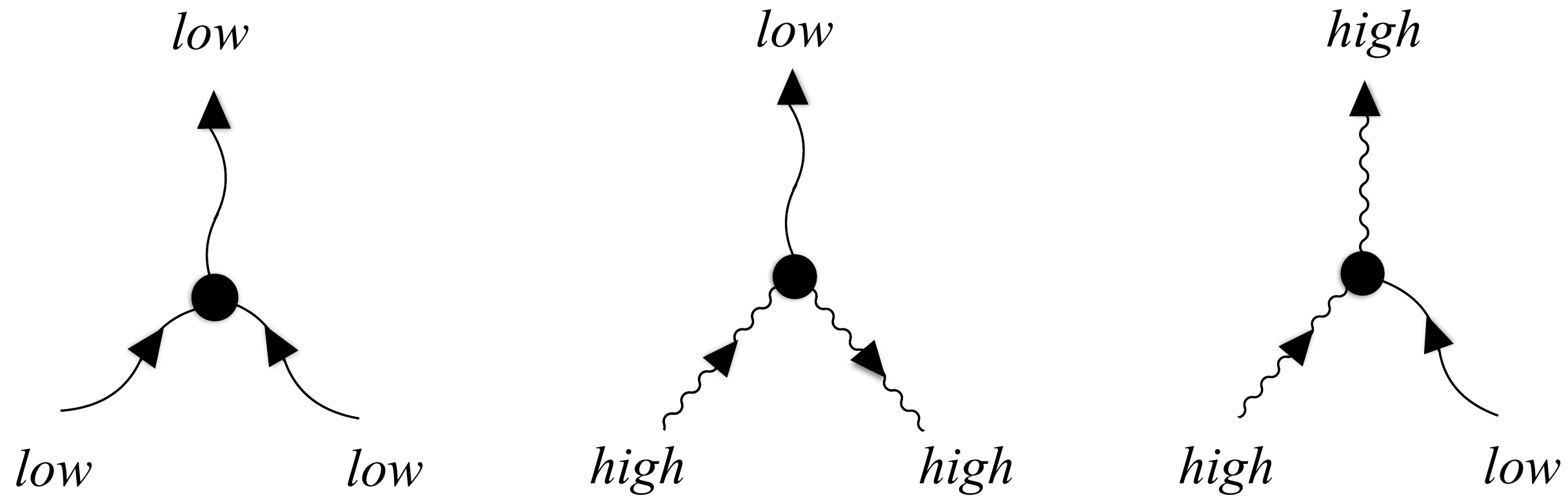}
  \caption{Set of 3 triadic interactions retained in the GQL approximation, out of 6 total.  The other 3 triads are discarded \citep{mct16}. Note that outgoing arrows are identical to incoming arrows with negative wavenumber}
  \label{fig:GQL}
 \end{figure}

We consider rotating plane Couette flow between two parallel plates as shown in Figure~\ref{fig:setup}. The parallel plates at $z=\pm h$ translate with velocity $\pm U_0$ in the $x$-direction, and the system rotates about the $y$-axis with rotation rate $\Omega$. The flow is described by the non-dimensional three-dimensional 
incompressible Navier-Stokes equations, \smtr{where we non-dimensionalise lengths with half the gap-width $h$ and times with $h/U_0$. This yields}\
\begin{equation}
 \pd_t \bm u + (\bm u \cdot \bm \nabla) \bm u +  \frac{1}{Ro}{\hat {\bm y}} \times \bm u = - \smtr{\bnabla} p + \frac{1}{Re} \grad^2 \bm u,
 \label{eq:ns}
\end{equation}
where $\bm u$ is the velocity field, \smtr{$p$} is the pressure and the non-dimensional parameters are the Reynolds number $Re = U_0 h/\nu$ and the Rossby number $Ro = U_0/(2 \Omega h)$\footnote{Note that many authors discussing RPC flow utilise the non-dimensional ``Rotation Number" which is the inverse Rossby number (and also termed $Ro$ just to add to the confusion).}. No-slip boundary conditions are applied at $z=\pm 1$ in non-dimensional co-ordinates --- so we set
\begin{equation}
\bm u = (u,v,w) = (\pm1,0,0) \,\, {\rm at} \,\,z=\pm \smtr{1},
\label{NSbc}
\end{equation}
whilst the domain is assumed to be periodic in the two horizontal directions at $x=0, L_x$ and $y=0, L_y$. 
Equation~(\ref{eq:ns}) together with boundary conditions~(\ref{NSbc}) are solved using a pseudo-spectral code, which employs Fourier expansions in the two horizontal periodic directions and uses Chebychev expansions in the $z$-direction.  Timestepping is performed via a third-order semi-implicit \smt{backward differentiation formula} (BDF) scheme  --- the code utilises the flexible {\it Dedalus} PDE solving framework \citep{dedalus}. We use a resolution of $128 \times 128 \times 64$ for the two lower Reynolds numbers and $256 \times 256 \times 128$ for the $Re=3000$ case. \smt{These numerical resolutions were verified as sufficient by comparing with shorter runs at higher resolution.} All calculations are run until a statistically steady state is reached.

\subsection{\label{gql}The GQL approximation}
\begin{figure}
\centerline{\includegraphics[width=5in]{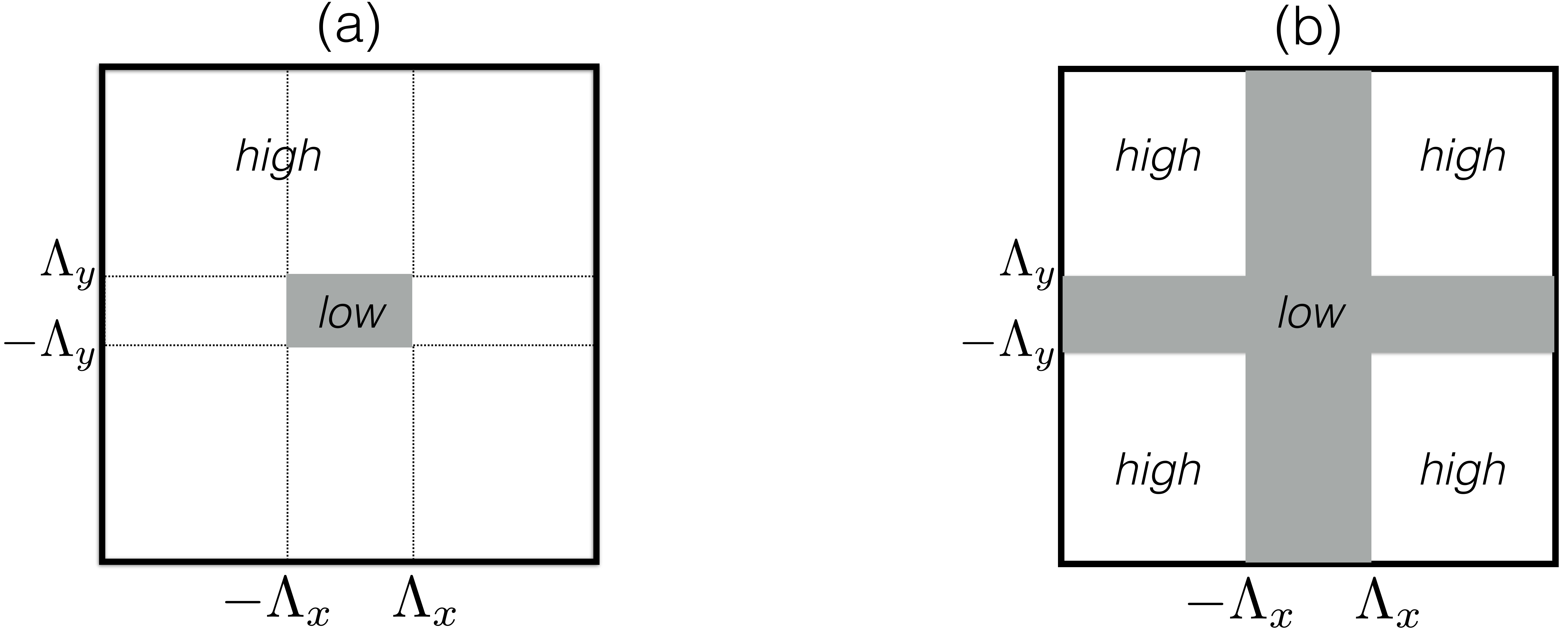}}
\caption{Separation in wavevector space of modes into `low' and `high' components.  (a) The separation employed in this paper.  (b) An alternative separation scheme.}
\label{fig:HighLow}
\end{figure}

The Generalised Quasilinear Approximation was introduced by \citet{mct16} to extend the often-utilised quasilinear approximation to include self-consistent interactions of large-scale modes; also see \cite{chmt16} and \cite{Constantinou:2016fp}.  Here we extend the approximation to the three-dimensional case for which there is translational symmetry in 2 of the 3 dimensions. Specifically we perform a idempotent decomposition of the velocity and pressure into large-scale and small-scale modes, i.e.\ we set, say, $\bm u(x, y, z)= \bm u_l + \bm u_h$, where
\jbmr{\begin{equation}
\bm u_l(x, y, z)
=\sum_{k_x=-\Lambda_x}^{\Lambda_x}\sum_{k_y=-\Lambda_y}^{\Lambda_y}{\bm u}_{\bf k}(z) \,e^{i  {2 \pi k_x x}/{L_x}+
i {2 \pi k_y y}/{L_y}}, \quad \quad
%+ \sum_{k=\Lambda_x+1}^{N_{k_x}}\sum_{l=\Lambda_y+1}^{N_{k_y}}{\bm u}_{kl}(z) \,e^{ik^\prime x+il^\prime y},
\bm u_h = \bm u - \bm u_l, 
\end{equation}
where $\bm u_l$ and $\bm u_h$ are the `low' and `high' wavenumber modes respectively.   \smt{Similarly $p(x,y,z) = p_l+p_h$.}}
The separation between high and low components in wavevector space is shown in Figure \ref{fig:HighLow}a.  We note here that when $\Lambda_x = \Lambda_y = 0$ the low modes are the horizontally averaged (mean) modes and the high modes are the fluctuations about that mean (sometimes termed eddies).  
Other divisions into `low' and `high' are possible; see for instance Figure \ref{fig:HighLow}b.  Here we adopt the choice of Figure \ref{fig:HighLow}a as, for this choice, no modes with short lengthscales (and consequently small \smt{timescales}) are included in the low modes. This is useful as the  statistics of the high modes may then be viewed as a subgrid model for the low modes, and there is no mixture of timescales between the evolving low-modes and the subgrid high modes. 

\smtr{The QL approximation is then characterised by the suppression of certain mode interactions in  the dynamics. In particular, though  mean/eddy $\rightarrow$ eddy and eddy/eddy $\rightarrow$ mean interactions are kept, eddy/eddy $\rightarrow$ eddy interactions are discarded \citep[\smtr{as for example done in }][]{herring1963,lh1968,dickinson1969,tobias2011astrophysical}. When $\Lambda_x, \Lambda_y  \ne 0$ the Generalised Quasilinear approximation differs from QL ---  the triad interactions that are retained and discarded are selected so as to enable closure and preserve the relevant conservation laws  \citep{mct16}. 
Here the interactions  low/low $\rightarrow$ low, high/high $\rightarrow$ low and low/high $\rightarrow$ high are retained with all other interactions discarded. \jbmr{Again, retaining only this set of interactions, with both $\Lambda_x=0$ and $\Lambda_y=0$, is QL.}  If one of $\Lambda_x$ or $\Lambda_y=0$ then the system is Quasilinear in that direction, but may include more triad interactions in the other direction. Furthermore as $\Lambda_x, \Lambda_y \rightarrow \infty$ the GQL system consists solely of fully interacting low modes and returns to fully nonlinear Direct Numerical Simulation.   Finally if the system has one of $\Lambda_x$ or $\Lambda_y=0$ with the other cut-off set so that all the modes are retained then the system is Quasilinear in one direction of symmetry and fully nonlinear in the other; the RNL systems described in the introduction with $\Lambda_x=0$ and all modes kept in the $y$-direction are in this class of approximation. Thus GQL is a controlled approximation that systematically improves upon QL.}

\smt{To be clear, when the GQL approximations is implemented numerically the same spectral resolution is employed, but certain triad interactions are suppressed in the streamwise and spanwise directions (all modes and interactions are retained in the $z$-direction). There are a number of ways of suppressing the nonlinear interactions in a spectral code, however we choose to derive separate equation sets to be satisfied by the low and high modes; these take the form
\begin{eqnarray}
 \pd_t \bm u_l + \langle (\bm u_l \cdot \bm \nabla) \bm u_l +  (\bm u_h \cdot \bm \nabla) \bm u_h \rangle_l+  \frac{1}{Ro}{\hat {\bm y}} \times \bm u_l &=& - \smtr{\bnabla} p_l + \frac{1}{Re} \grad^2 \bm u_l,\\
 \pd_t \bm u_h + \langle (\bm u_l \cdot \bm \nabla) \bm u_h +  (\bm u_h \cdot \bm \nabla) \bm u_l \rangle_h+  \frac{1}{Ro}{\hat {\bm y}} \times \bm u_h &=& - \smtr{\bnabla} p_h + \frac{1}{Re} \grad^2 \bm u_h,
\label{eq:nslh}
\end{eqnarray}
where $\langle \rangle_l$ and $\langle \rangle_h$ represent the projection of the nonlinear terms onto the low and high wavenumbers respectively. We note that this is not the most efficient way of implementing GQL and we stress that this formalism does not therefore currently offer any saving over DNS. Rather we are using this approach to determine the effectiveness of the GQL approximation. If GQL can be demonstrated to yield and accurate description of the statistics then a statistical description of the high modes via Direct Statistical Simulation will offer a significant reduction in computational cost.}

\section{Results}
\label{sec:results}

\subsection{DNS}
We begin the results section by describing the behaviour of the RPC system as revealed by DNS. For this geometry the non-dimensional mean absolute vorticity is given by \smtr{$1/Ro + d{\overline U}/dz$ (where $U(z)$ is the mean flow)} and so we expect the instability to be enhanced for negative Rossby number and suppressed for positive $Ro$ (we note in this co-ordinate system this is opposite to that for \citet{ba96} and subsequent papers). Much of the nonlinear behaviour can --- and has --- been interpreted in terms of turbulence and secondary flows mixing the mean absolute vorticity in the core of the flow to suppress the instability.  For a thorough discussion of the competition between turbulent transport and that provided by the  secondary flows see \citet{ba97,suryadi14} and the references therein.

\begin{figure}
\begin{subfigure}{0.50\textwidth}
\caption{$Re=1300$  $Ro=-100$}
   \includegraphics[width=\textwidth]{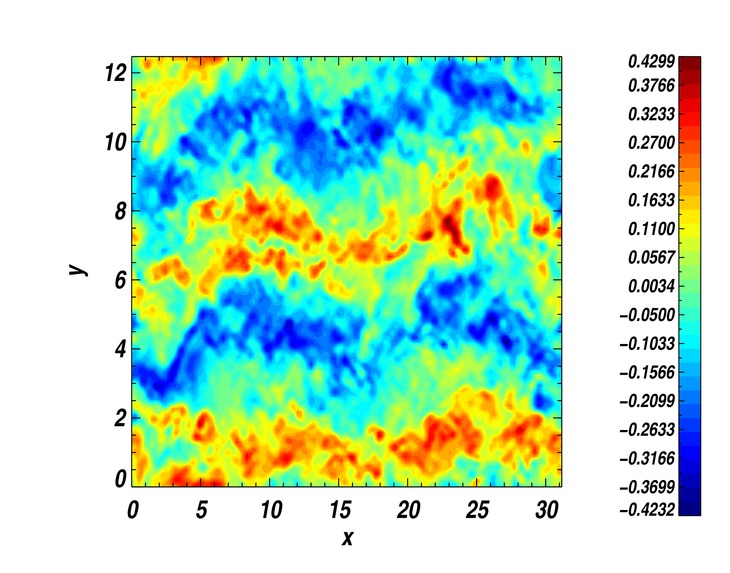}
 %\caption{}
   \label{fig:u_weak}
\end{subfigure}
 \begin{subfigure}{0.5\textwidth}
\caption{$Re=1300$  $Ro=-10$}
   \includegraphics[width=\textwidth]{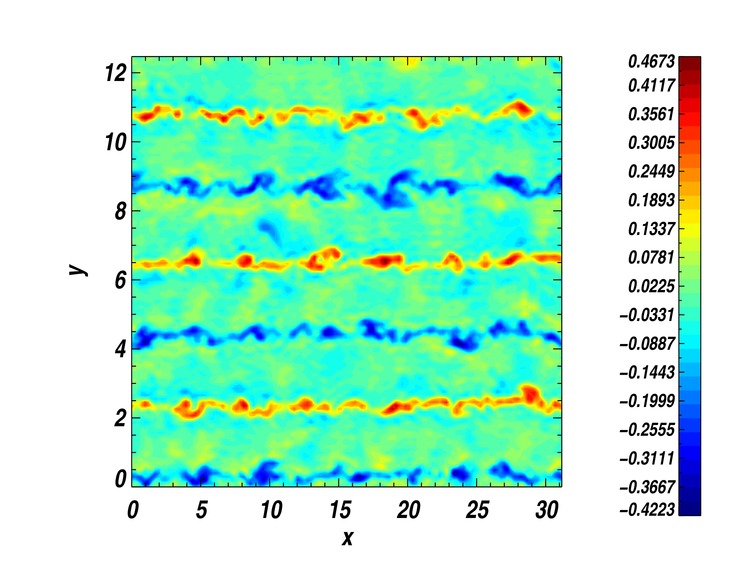}
%   \caption{}
   \label{fig:helb}
 \end{subfigure}
%\begin{subfigure}{0.240\textwidth}
%   \includegraphics[width=\textwidth]{./new_f4c.jpg}
%   \caption{}
%   \label{fig:u_mean}
%\end{subfigure}
 \begin{subfigure}{0.5\textwidth}
\caption{$Re=3000$ $Ro=-100$}
   \includegraphics[width=\textwidth]{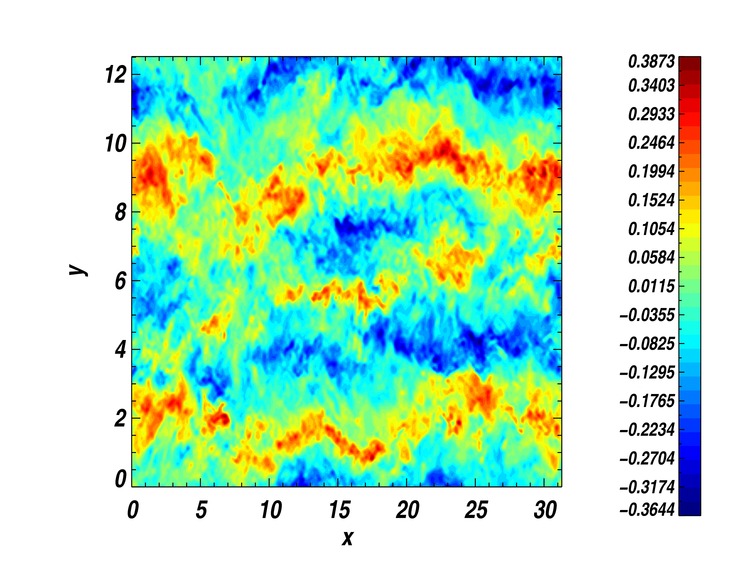}
%   \caption{{\bf CHANGE THIS ONE}}
   \label{fig:helc}
 \end{subfigure}
  \caption{(Colour online) Instantaneous snapshots of the streamwise velocity $u(x,y)$ at $z=0.1$ for $Re=1300$ and (a) $Ro=-100$, (b) $Ro=-10$, and (c) $Re=3000$, $Ro=-100$.}
  \label{fig:dns}
 \end{figure}

\begin{figure}
\begin{subfigure}{0.50\textwidth}
\caption{}
   \includegraphics[width=\textwidth]{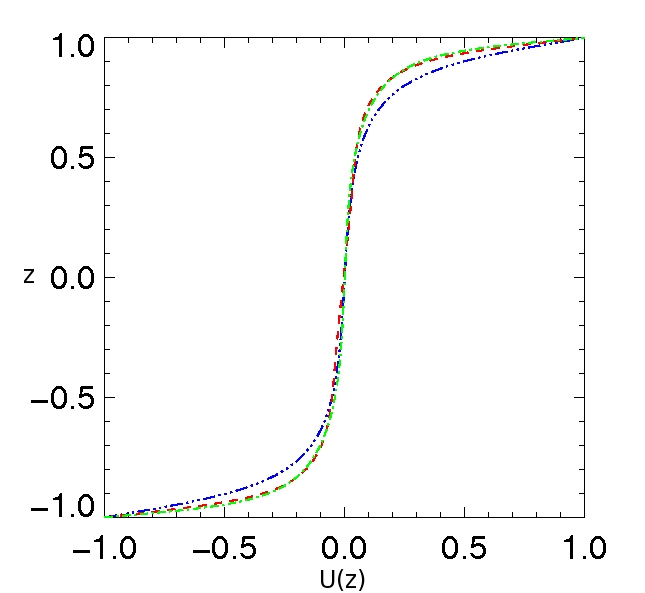}
 %\caption{}
%   \label{fig:u_weak}
\end{subfigure}
 \begin{subfigure}{0.5\textwidth}
\caption{}
   \includegraphics[width=\textwidth]{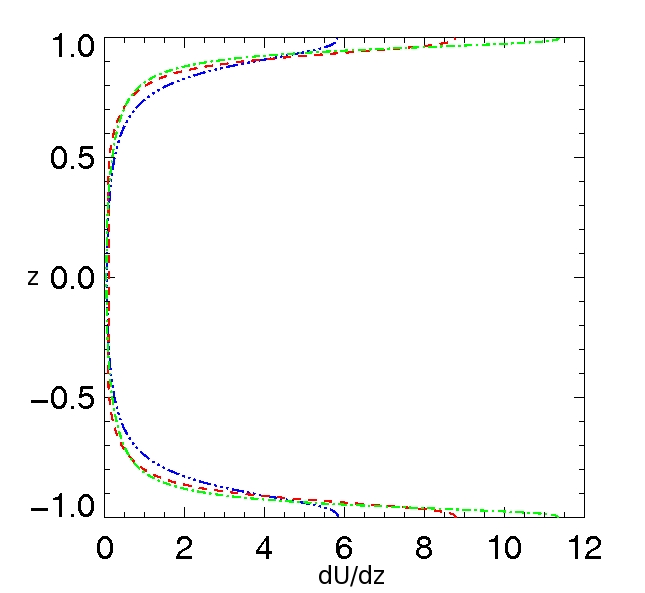}
%   \caption{}
 %  \label{fig:helb}
 \end{subfigure}
%\begin{subfigure}{0.240\textwidth}
%   \includegraphics[width=\textwidth]{./new_f4c.jpg}
%   \caption{}
%   \label{fig:u_mean}
%\end{subfigure}
\caption{(Colour online) (a) Mean flows U(z) and (b) Mean shears $dU/dz$ as a function of $z$ (vertical axis) for $Re=1300$, $Ro=-100$ (blue, dot dot dashed), $Re=1300$, $Ro=-10$ (red dashed),  and  $Re=3000$, $Ro=-100$ (green \smtr{dot-dashed}).\smt{We note that the red and green curves lie nearly on top of each other.}}
\label{fig:u_du_dns}
 \end{figure}

In this paper, we set $L_x=10 \pi$, $L_y=4 \pi$, and vary  $Ro$ and $Re$ as shown in Table~1. The fiducial run is weakly rotating $Ro=-100$ and at moderate $Re=1300$.  For this choice of parameters the solution is particularly complicated and so presents a significant test for GQL. The solution has three constituent parts; a mean flow, a secondary flow and a turbulent part \citep{ba96}. The secondary flow takes the form of (initially three and then two) time-dependent turbulent sinuous rolls largely aligned with the $x$-axis (as shown in Figure~\ref{fig:dns}a ). 
\smt{Figure~\ref{fig:u_du_dns} shows the mean flow $U(z)$ and mean shear $dU/dz$ as a function of depth. Here the averaging has been performed over horizontal planes {\it and} the last third of the calculation ($500 \le t \le 750$); \smtr{the solutions are run up to $t=750$ from rest}. All statistical quantities such as spectra and second cumulants are also averaged in time over this interval.} 
These rolls and the anisotropic turbulence act so as to redistribute the vorticity, forming a core with a weak shear with two matching layers of strong shear (see the blue line of Figure~\ref{fig:u_du_dns}). The statistically steady state is achieved with the energy input through the boundaries being matched on average by the viscous dissipation, given by $D_\nu = \frac{1}{Re}{\int \omega^2 \,dV}$; we shall use this dissipation rate as a measure to evaluate the effectiveness of the GQL approximation (see Table~\ref{tbl:param}). For moderate rotation ($Ro=-10.0$) the solution is less turbulent and the (less time-dependent) secondary flow has three relatively straight rolls (Figure~\ref{fig:dns}b). These are more efficient at transporting the mean flow (green line of Figure~\ref{fig:u_du_dns}) and so the core has a gradient closer to zero with steeper shears in the wings. 
Finally we note that for more turbulent, weakly rotating flow ($Re=3000$, $Ro=-100$) both the secondary flow, which again has (on average two) rolls largely aligned with the $x$-axis (see Figure~\ref{fig:dns}c) and the turbulence are acting strongly to transport the momentum (red lines of Figure~\ref{fig:u_du_dns}).
\begin{table}
  \begin{center}
\def~{\hphantom{0}}
    \begin{tabular}{c*{16}{c}}
Case&{Ro}   & {Re}   & {\bf all/all} &$0/0$  & $0$/all  & all/$0$ & $3/3$ &  $3/10$ & $5/5$ & $6/6$ \\
\hline     
A&{$-100$} & $1300$ &{\bf 11.6} &16.0&10.4  &11.7 & 11.6  &11.4  &11.1&11.0 \\
B&{$-10$} & $1300$ &{\bf 17.7}     &29.4&22.8  &20.7  &18.6  &17.2  &17.8 & 16.6 \\
%{$-2$} & $1300$     &{\bf 14.8}& 23.7 &18.2  &18.66  &  & & && \\
C&{$-100$} & $3000$ &  {\bf 22.7}   &26.9&16.3  &18.1  &19.2  &18.2  &19.5 &19.5  \\

   \end{tabular}
  \caption{Enstrophy $\int_V \omega^2 \, dV$ as a function of cut-off wavenumbers $\Lambda_x$ and $\Lambda_y$.} 
  \label{tbl:param}
   \end{center}
\end{table}

We conclude this section by noting that, as discussed by \cite{diamondetal05}, a measure of the departure from quasilinearity in turbulent fluid systems is often given by the Kubo number $R = u_{rms} \tau_c / l_c$ where $\tau_c$ is the correlation time of the turbulence and $l_c$ its correlation length. \smt{The Kubo number is therefore a measure of the ratio of the  advective (nonlinear) term to the time-derivative term in the equation for the fluctuating velocity. For short correlation time turbulence $R$ is asymptotically small, the nonlinear term may be formally neglected and the system is well represented by a quasilinear approximation. We calculate the correlation time using the first zero of the temporal auto-correlation function and the correlation length via a typical lengthscale of variation of the two-point (in space) correlation function.} For these flows we note that $R \sim {\cal O}(0.5)$, with the Kubo number being smallest for Case B. \smt{Thus for these systems there is no a priori reason why the quasilinear approximation should hold.}

%\begin{figure}
%\centerline{\includegraphics[width=3.5in]{./fig5.pdf}}
%\caption{(Colour online) (a) Mean flows U(z) and (b) Mean shears $dU/dz$ as a function of $z$ (vertical axis) for $Re=1300$, $Ro=-100$ (blue), $Re=1300$, $Ro=-10$ (red),  and  $Re=3000$, $Ro=-100$ (green).}
%\label{fig:u_du_dns}
%\end{figure}

\subsection{Evaluation of the GQL approximation}

\begin{figure}
 \centering
 \begin{subfigure}{0.40\textwidth}
 \caption{$\Lambda_x=64$  $\Lambda_y=64$}
   \includegraphics[width=\textwidth]{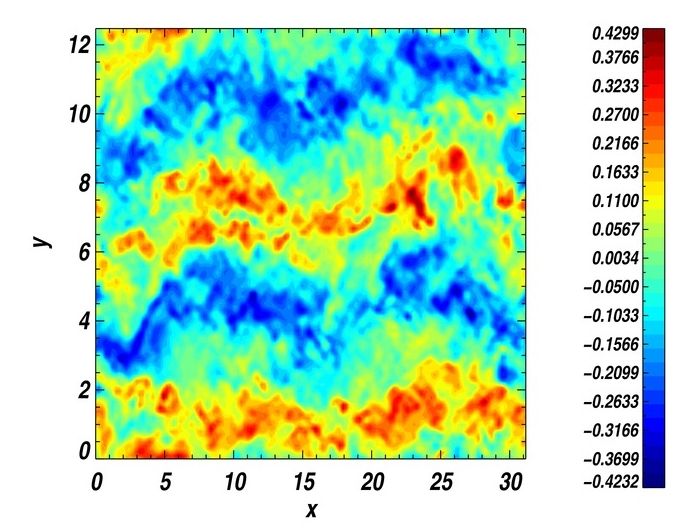}
   \label{fig:sn_64_64_weak}
\end{subfigure}
 \begin{subfigure}{0.40\textwidth}
   \caption{$\Lambda_x=0$ $\Lambda_y=0$} 
  \includegraphics[width=\textwidth]{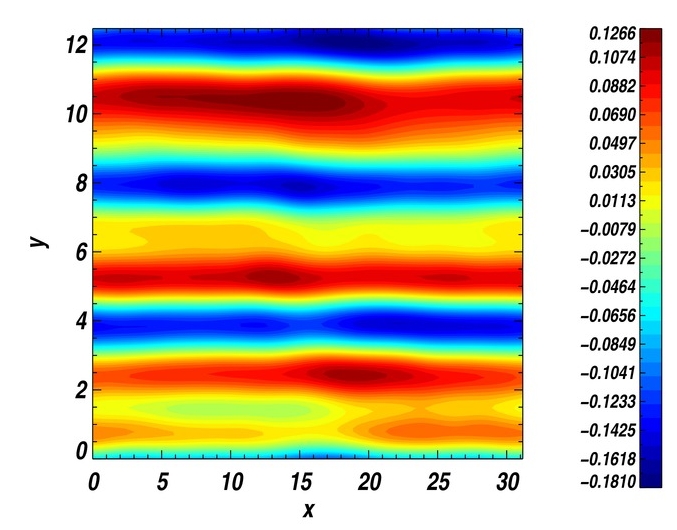}
   \label{fig:sn_0_0_weak}
 \end{subfigure}
\begin{subfigure}{0.40\textwidth}
   \caption{$\Lambda_x=0$ $\Lambda_y=64$}
   \includegraphics[width=\textwidth]{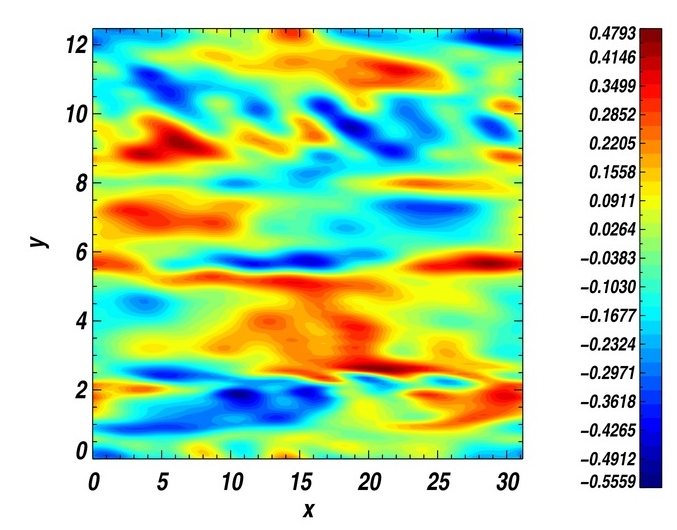}
   \label{fig:sn_0_64_weak}
\end{subfigure}
 \begin{subfigure}{0.40\textwidth}
   \caption{$\Lambda_x=64$ $\Lambda_y=0$}
   \includegraphics[width=\textwidth]{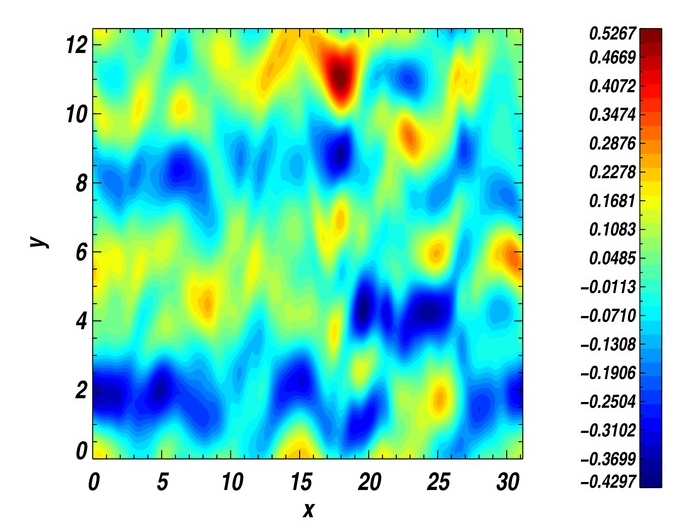}
   \label{fig:sn_64_0_weak}
 \end{subfigure}
\begin{subfigure}{0.40\textwidth}
   \caption{$\Lambda_x=5$ $\Lambda_y=5$}
   \includegraphics[width=\textwidth]{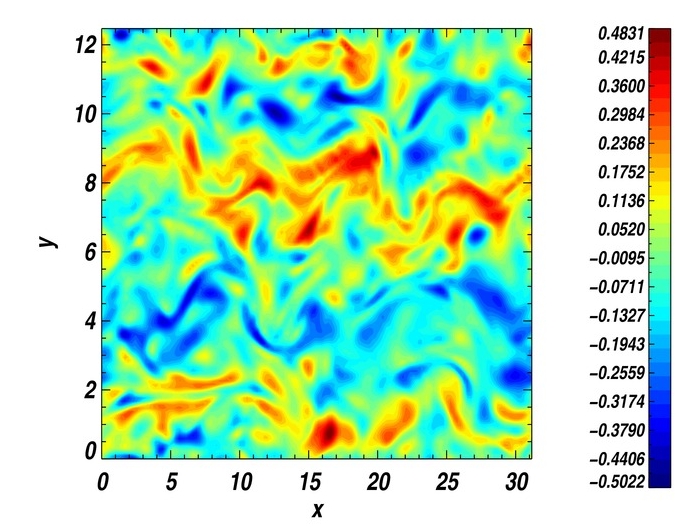}
   \label{fig:sn_5_5_weak}
\end{subfigure}
 \begin{subfigure}{0.40\textwidth}
   \caption{$\Lambda_x=3$ $\Lambda_y=10$}
   \includegraphics[width=\textwidth]{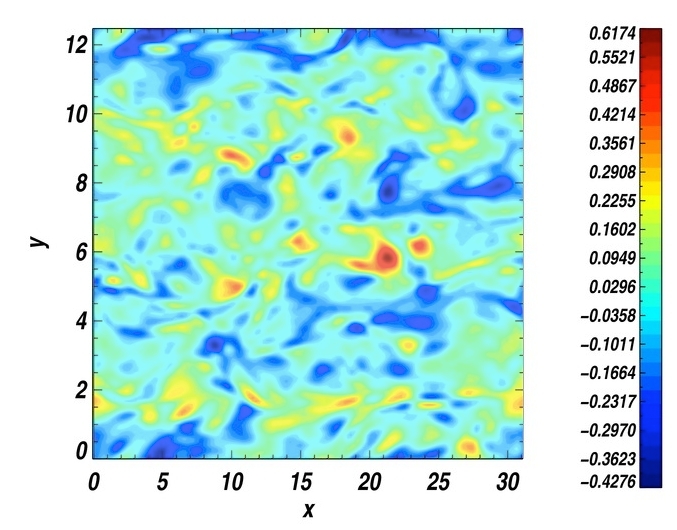}
   \label{fig:sn_3_10_weak}
 \end{subfigure}
\begin{subfigure}{0.40\textwidth}
   \caption{$\Lambda_x=3$ $\Lambda_y=3$}
   \includegraphics[width=\textwidth]{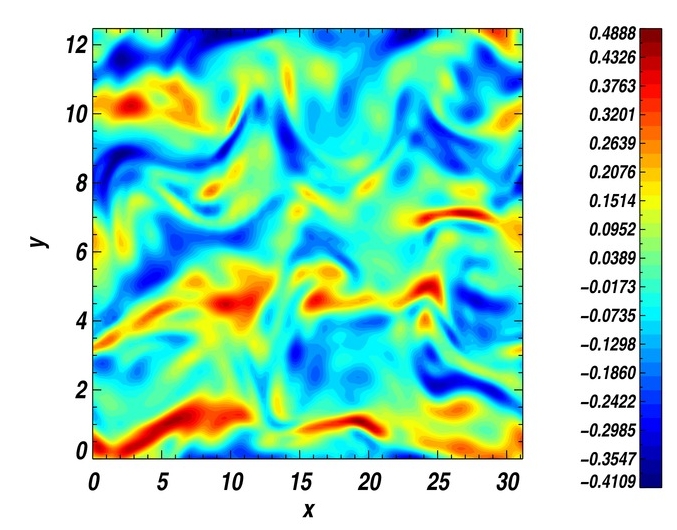}
   \label{fig:sn_3_3_weak}
\end{subfigure}
 \begin{subfigure}{0.40\textwidth}
   \caption{$\Lambda_x=10$ $\Lambda_y=3$}
   \includegraphics[width=\textwidth]{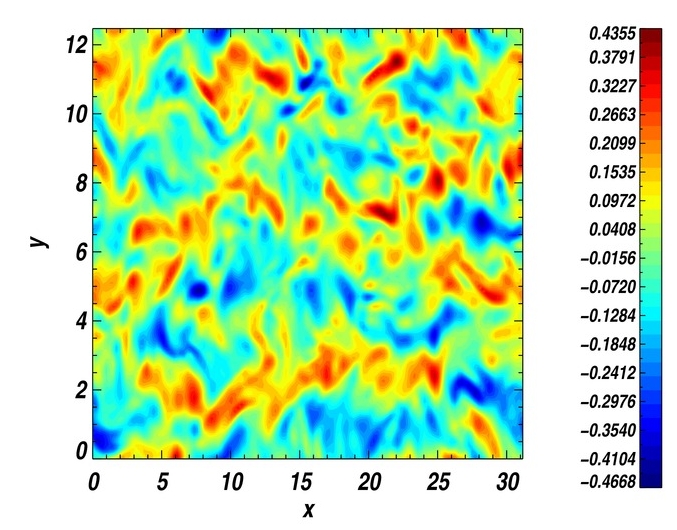}
   \label{fig:sn_6_6_weak}
 \end{subfigure}
% \begin{subfigure}{0.4\textwidth}
%   \caption{}
%   \includegraphics[width=\textwidth]{./sp_6_6_weak.jpg}
%   \label{fig:sp_6_6_weak}
% \end{subfigure} 
%\begin{subfigure}{0.4\textwidth}
%   \caption{}
%   \includegraphics[width=\textwidth]{./sp_6_6_weak.jpg}
%   \label{fig:sp_6_6_weak}
% \end{subfigure}
%\vskip -0.2truein 
%\begin{subfigure}{0.58\textwidth}
%   \includegraphics[width=\textwidth]{./sp_colbar.jpg}
%   \caption{}
%   \label{fig:sp_colbar}
% \end{subfigure}
  \caption{\smt{(Colour online) Instantaneous snapshots of the evolution of the flow for case A. The snapshots show the streamwise velocity $u(x,y)$ at $z=0.1$ and $t=750$.}  \jbmr{The flows are intermittent; therefore it is more meaningful to compare statistics than snapshots.}}
 \end{figure}

Having established the dynamics of the RPC system via DNS,  we evaluate the GQL approximation for various truncations $(\Lambda_x,\Lambda_y)$ in the streamwise and spanwise directions. Clearly the presence of mean flows and rotation make the turbulence highly anisotropic and so there is no {\it a priori} reason why $\Lambda_x$ should be set equal to $\Lambda_y$. This yields a huge range of choice for these cut-offs and we present only  a representative few below. Previous evaluations of GQL have only considered two-dimensional models, with only one direction of splitting into low and high modes. We shall evaluate the degree of success of the approximation by determining how well they reproduce the low order statistics of the system; mean flows, two-dimensional spectra, two-point correlation functions and total enstrophy. 

 \begin{figure}
 \centering
% \begin{subfigure}{0.40\textwidth}
%   \includegraphics[width=\textwidth]{./u_new.jpg}
%   \caption{}
%   \label{fig:u_mean}
%\end{subfigure}
 \begin{subfigure}{0.45\textwidth}
   \caption{$Re=1300$ $Ro=-100$}
\vskip-0.25cm
   \includegraphics[width=\textwidth]{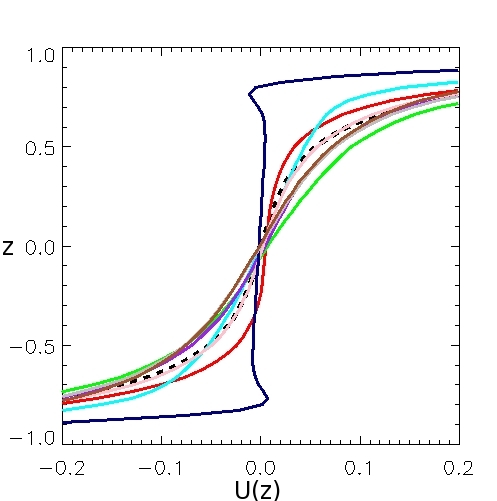}
   \label{fig:tmp_weak_new}
 \end{subfigure}
 \begin{subfigure}{0.45\textwidth}
   \caption{$Re=1300$ $Ro=-10$}
\vskip-0.25cm
   \includegraphics[width=\textwidth]{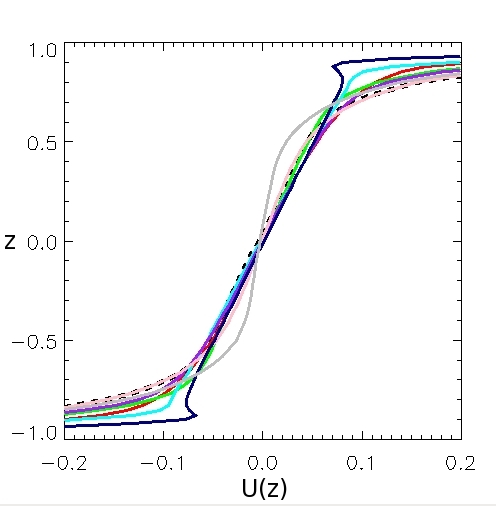}
   \label{fig:tmp_mod_new}
 \end{subfigure}
 \begin{subfigure}{0.45\textwidth}
   \caption{$Re=3000$ $Ro=-100$}
\vskip-0.25cm
   \includegraphics[width=\textwidth]{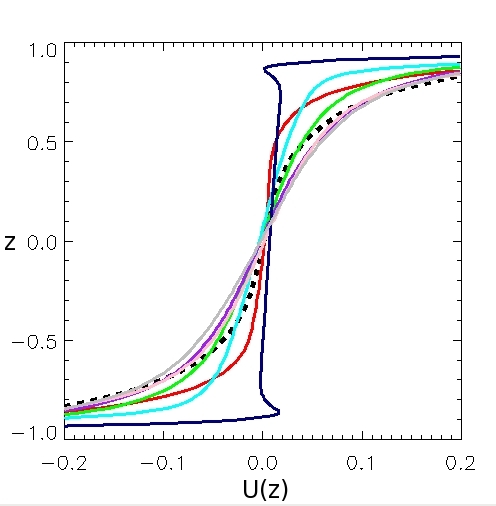}
   \label{fig:tmp_mod_re_new}
 \end{subfigure}
\caption{(Colour online) Mean flow $U(z)$  as a function of $z$ (vertical axis) for $Re=1300$, (a) $Ro=-100$,
(b) $Ro=-10.0$, (c) $Re=3000$, $Ro=-100$ Here DNS (dashed black line), $(\Lambda_x,\Lambda_y)=(0,0)$ (navy), $(0,64)$ (red), $(64,0)$ (cyan), $(3,3)$ (green), $(5,5)$ (purple), $(3,10)$ (pink), $(6,6)$ (grey), \smtr{$(10,3)$ (brown --- top panel only)}. Note  only $-0.2 \le U(z) \le 0.2$ is shown to give a close up of the region of reduced shear.}
\label{fig:first_comp}
\end{figure}

\smt{Although we are evaluating the effectiveness of GQL in describing the statistics of the rotating Couette system it is of interest to determine how the varying degrees of approximation perform in reproducing some of the dynamics of the system. In particular, for case A the system evolves significantly in time and produces coherent structures in the form of rolls largely aligned with the $x$-axis. At short time the solution has three largely straight rolls, which can be understood via the linear instability of the modified shear flow \citep{ba97}. For longer times these rolls evolve nonlinearly, interacting with the turbulence until the coherent rolls take the form of two wavy rolls as shown in Figure~\ref{fig:sn_64_64_weak}, which is a snapshot of the streamwise velocity (measured on the plane $z=0.1$) at the end of the simulation ($t=750$). Figures~6b-h show how well severe truncations of GQL perform in capturing the dynamics of the system. The quasilinear approximation in Figure~\ref{fig:sn_0_0_weak} appears to capture the initial instability of the modified shear state to three rolls, but not the subsequent evolution to two wavy rolls. GQL performs better, even at severe truncations. We draw attention to the snapshots in Figures~(\ref{fig:sn_5_5_weak}-\ref{fig:sn_6_6_weak}) which suggest that these truncations are capable of capturing the long-time nonlinear evolution including the merging of the coherent structures.  This will be quantified by measuring the statistics of the flows.}

Figure~\ref{fig:first_comp} shows how well the various degrees of approximation perform in reproducing the mean flow (or first cumulant --- $U(z)$) giving a close-up of the core of the mean flow. For the weakly rotating case --- in  a --- the suppression of the shear is greatly over-estimated by the quasi-linear approximation and fairly well approximated by GQL at all truncations. Note though that GQL performs poorly when either one of $\Lambda_x$ or $\Lambda_y=0$, with the suppression of the shear in the core being overestimated, for reasons we shall discuss later. For the moderately rotating flows (case B, shown in Figure~\ref{fig:first_comp}b) the shear in the core of the flow is well reproduced by all the approximations, although that near the boundaries is significantly over-estimated by QL. We believe that the improvement over that for case A in all the methods arises because the increase in rotation leads to a suppression of the role of the turbulence and a diminishment of the variability of the secondary flows. The secondary flow rolls are relatively straight, are persistent, and their role in transporting momentum is more easily captured by all the approximations. For this flow, although the secondary flow is still present, it is more ordered. Finally Figure~\ref{fig:first_comp}c shows the mean flows for the more turbulent, weakly rotating case ($Re=3000$, $Ro =-100$) with the various degrees of approximation. Again here the QL approximations significantly overestimate the degree of mixing in the core, and consequently overestimates the shear in the boundary layers. This deficiency is again shared by the calculations that are quasilinear in either the streamwise or spanwise direction --- with the full GQL approximation calculations performing better.

A quadratic statistic that is
often of interest is the energy dissipation rate, which is proportional to the enstrophy. Table~1 gives the total enstrophy for the three runs at various degrees of approximation. The first thing to note is that the enstrophy increases with either increasing rotation rate or increased Reynolds number. This is to be expected as in both of these cases the small-scales are more prevalent than for the moderately turbulent and weakly rotating case (case A); see the discussion of the two-dimensional spectra to follow. The dissipation must, of course, in a statistically steady state balance the energy input through the boundaries, which is given by the wall shear stress. Reference back to Figure~5b shows how the wall shear stress increases from case A to B to C in the same ratio as the enstrophy in Table~1.
Flows under the QL approximation significantly overestimate the enstrophy; the overestimation of the mixing in the core of the flow leads to strong gradients at the boundaries. These in turn lead to greater wall shear stress required to satisfy the boundary conditions. This leads to a consequent increase in the dissipation via an increase in the mean enstrophy.  Moreover flows that are quasilinear in one direction (i.e.\ either $\Lambda_x=0$ or $\Lambda_y=0$) also tend to perform poorly, whilst  GQL runs with $\Lambda_x$, $\Lambda_y \ge 3$ perform better for all runs. \smtr{We note at this point that the approach to the DNS values of the enstrophy is non-monotonic with increase in $\Lambda_x$ and $\Lambda_y$; this is discussed further below.}

\begin{figure}
 \centering
 \begin{subfigure}{0.30\textwidth}
 \caption{}
   \includegraphics[width=\textwidth]{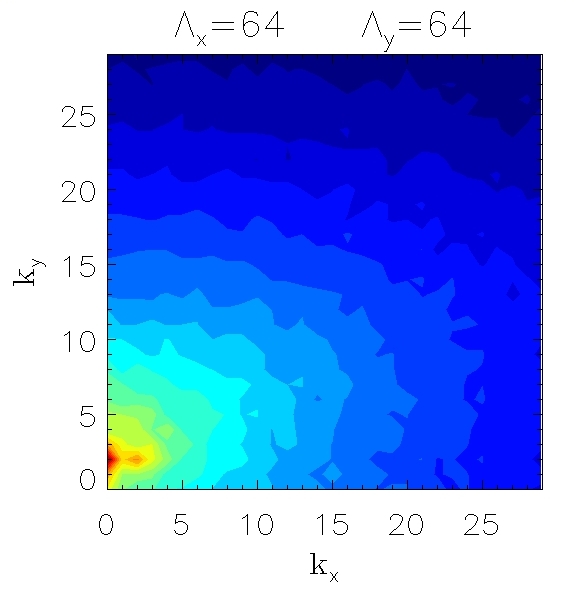}
   \label{fig:sp_64_64_weak}
\end{subfigure}
 \begin{subfigure}{0.30\textwidth}
   \caption{} 
  \includegraphics[width=\textwidth]{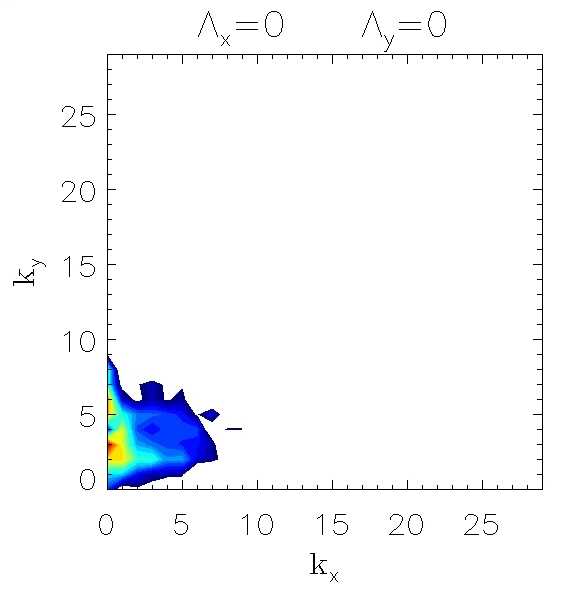}
   \label{fig:sp_0_0_weak}
 \end{subfigure}
\begin{subfigure}{0.30\textwidth}
   \caption{}
   \includegraphics[width=\textwidth]{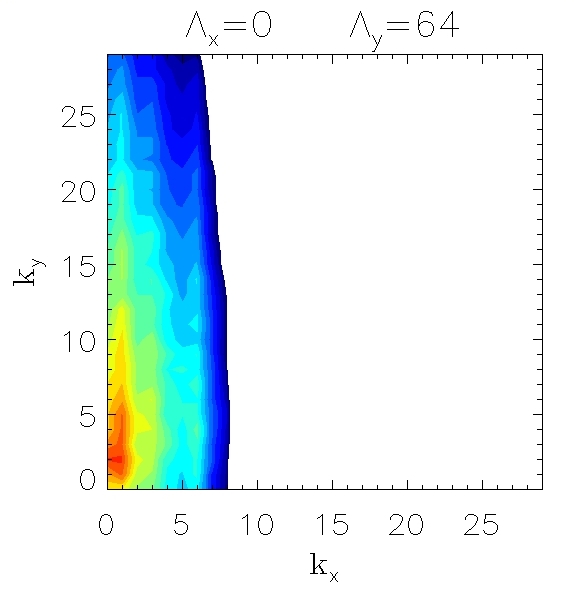}
   \label{fig:sp_0_64_weak}
\end{subfigure}
 \begin{subfigure}{0.30\textwidth}
   \caption{}
   \includegraphics[width=\textwidth]{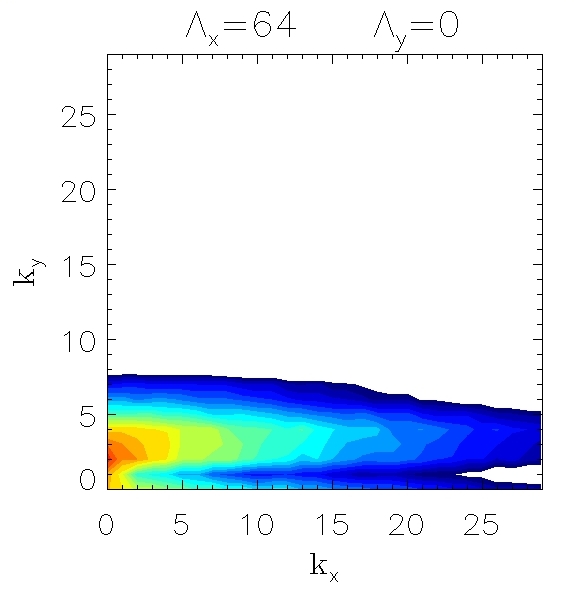}
   \label{fig:sp_64_0_weak}
 \end{subfigure}
\begin{subfigure}{0.30\textwidth}
   \caption{}
   \includegraphics[width=\textwidth]{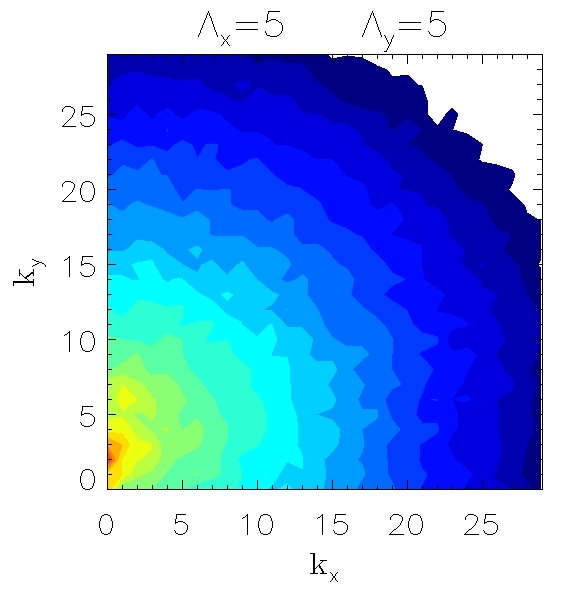}
   \label{fig:sp_5_5_weak}
\end{subfigure}
 \begin{subfigure}{0.30\textwidth}
   \caption{}
   \includegraphics[width=\textwidth]{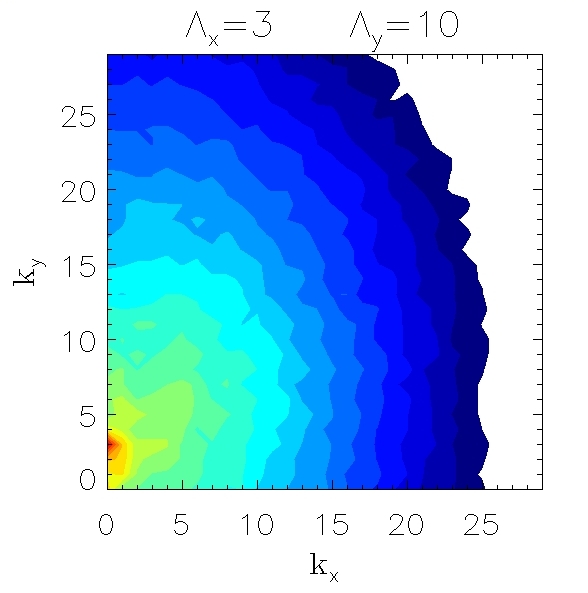}
   \label{fig:sp_3_10_weak}
 \end{subfigure}
\begin{subfigure}{0.30\textwidth}
   \caption{}
   \includegraphics[width=\textwidth]{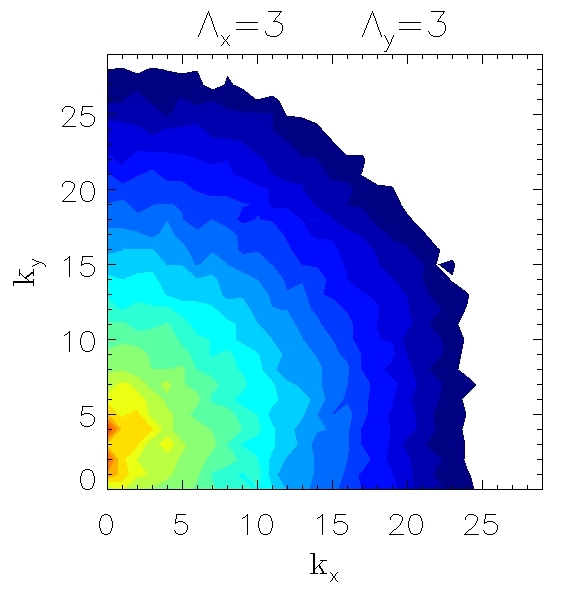}
   \label{fig:sp_3_3_weak}
\end{subfigure}
 \begin{subfigure}{0.305\textwidth}
   \caption{}
   \includegraphics[width=\textwidth]{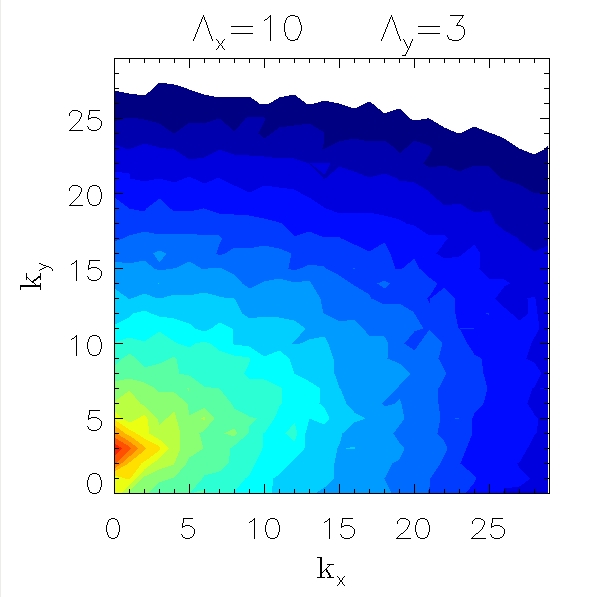}
   \label{fig:sp_6_6_weak}
 \end{subfigure}
% \begin{subfigure}{0.4\textwidth}
%   \caption{}
%   \includegraphics[width=\textwidth]{./sp_6_6_weak.jpg}
%   \label{fig:sp_6_6_weak}
% \end{subfigure} 
%\begin{subfigure}{0.4\textwidth}
%   \caption{}
%   \includegraphics[width=\textwidth]{./sp_6_6_weak.jpg}
%   \label{fig:sp_6_6_weak}
% \end{subfigure}
%\vskip -0.2truein 
\begin{subfigure}{0.5\textwidth}
   \includegraphics[width=\textwidth]{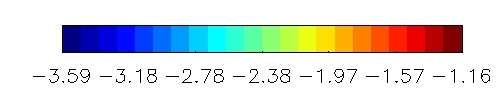}
%   \caption{}
   \label{fig:sp_colbar}
 \end{subfigure}
  \caption{(Colour online) Two-dimensional spectra for $u$ in the horizontal plane $z=0.1$. Case A is shown (Table \ref{tbl:param}).  The energy is shown on a \smt{(base 10)} logarithmic scale given by the colour-bar, with values below the minimum of the colour-bar not assigned a colour (so they appear white).}
  \label{fig:spec_weak}
 \end{figure}

The reason for this improvement of GQL over QL can be ascertained from an examination of the two-dimensional spectra of $u$; an  example for the DNS of case A is given in Figure~\ref{fig:spec_weak}a. These are calculated for a fixed horizontal plane $z=0.1$ just above the midplane and averaged over time. These show the anisotropy of the flow in the horizontal plane for the full solutions; the secondary flow apparent in the snapshot of Figure~\ref{fig:dns}a  dominates this spectrum at $(k_x,k_y) = (0,2)$ and $(2,2)$, though there is much energy in the surrounding wavenumbers. Energy cascades from the mean flows and secondary flows to smaller scales in an anisotropic fashion, with the energy falling more rapidly in the streamwise ($k_x$) direction than in the spanwise ($k_y$) direction. The shear inhibits the spectral transfer from large scales anisotropically, as expected.

The corresponding spectra for the QL and GQL approximations are shown in Figure~\ref{fig:spec_weak}b-h.  What is clear from Figure~\ref{fig:spec_weak}b, which shows the spectrum for QL,  is that  under this approximation the energy is concentrated at large scales  in both directions. Because of the quasilinear approximation, these wavenumbers (which of course are considered ``high") must be those to which energy feeds in directly from an instability of the mean flow. This instability may have occurred at any stage of the evolution to the statistically steady state. Once these high modes have reached finite amplitude they are sustained by the saturated shear and act to alter the shear flow via self-interaction (though of course some modes that arise from an instability of an intermediate shear  may not be sustained by the final shear profile --- and these will have decayed away. The ``high'' modes do not interact with each other though and so are not able to redistribute energy among themselves. It is clear therefore that the mean flow is not directly unstable to modes at small spatial scales and so there is no mechanism for these modes to receive energy.

\begin{figure}
 \centering
 \begin{subfigure}{0.320\textwidth}
\caption{}
   \includegraphics[width=\textwidth]{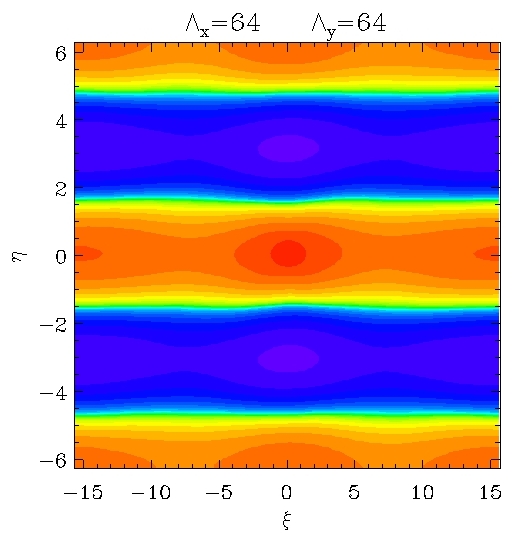}
 %\caption{}
   \label{fig:tmp_uu_64_64}
\end{subfigure}
 \begin{subfigure}{0.32\textwidth}
\caption{}   \includegraphics[width=\textwidth]{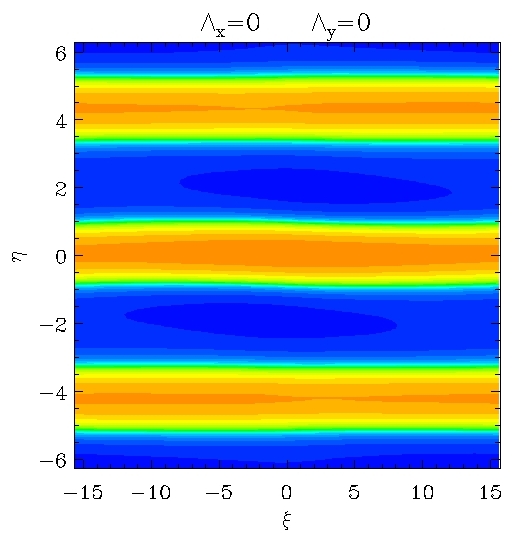}
%   \caption{}
   \label{fig:tmp_uu_0_0}
 \end{subfigure}
\begin{subfigure}{0.32\textwidth}
\caption{}   \includegraphics[width=\textwidth]{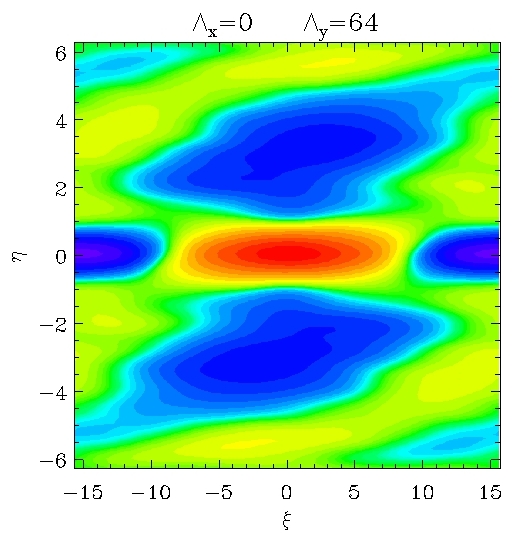}
%   \caption{}
   \label{fig:tmp_uu_0_64}
\end{subfigure}
 \begin{subfigure}{0.32\textwidth}
\caption{}   \includegraphics[width=\textwidth]{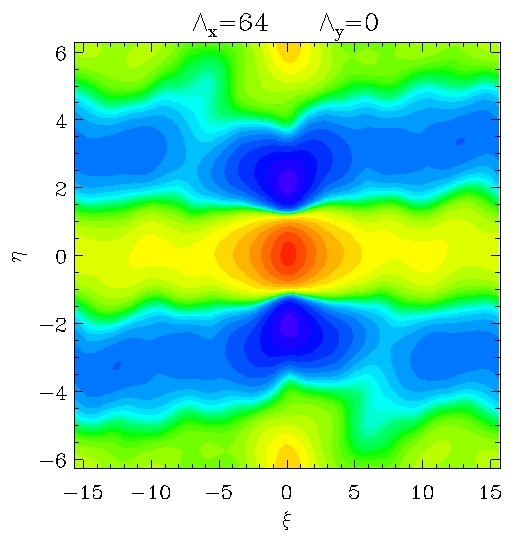}
%   \caption{}
   \label{fig:tmp_uu_640}
 \end{subfigure}
\begin{subfigure}{0.320\textwidth}
\caption{}   \includegraphics[width=\textwidth]{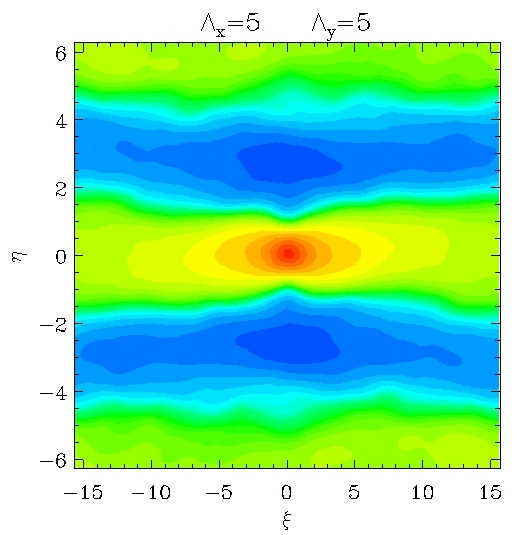}
%   \caption{}
   \label{fig:tmp_uu_55}
\end{subfigure}
 \begin{subfigure}{0.32\textwidth}
\caption{}   \includegraphics[width=\textwidth]{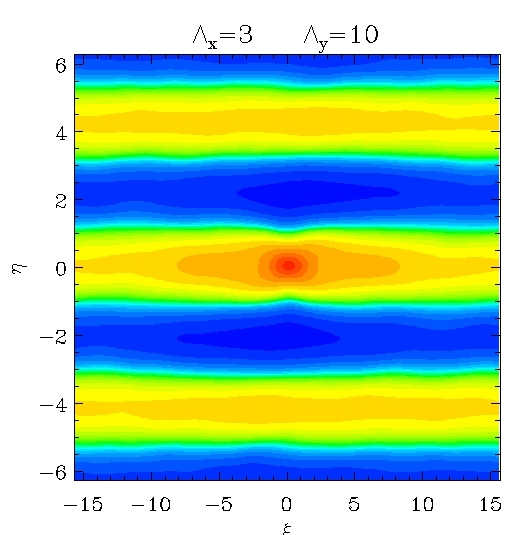}
%   \caption{}
   \label{fig:tmp_uu_3_10}
 \end{subfigure}
\begin{subfigure}{0.32\textwidth}
\caption{}   \includegraphics[width=\textwidth]{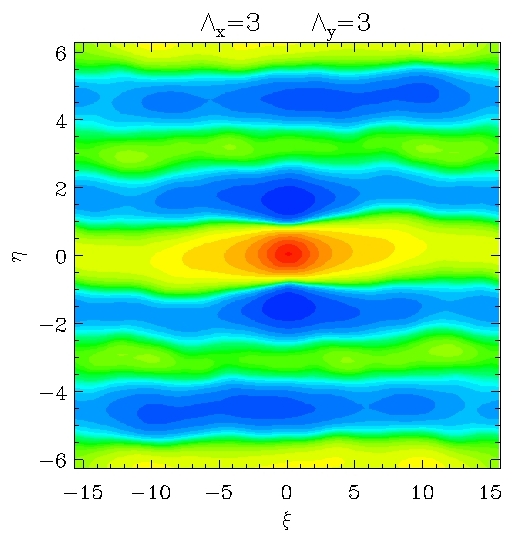}
%   \caption{}
   \label{fig:tmp_uu_3_3}
\end{subfigure}
 \begin{subfigure}{0.32\textwidth}
\vskip0.1cm
\caption{}   \includegraphics[width=\textwidth]{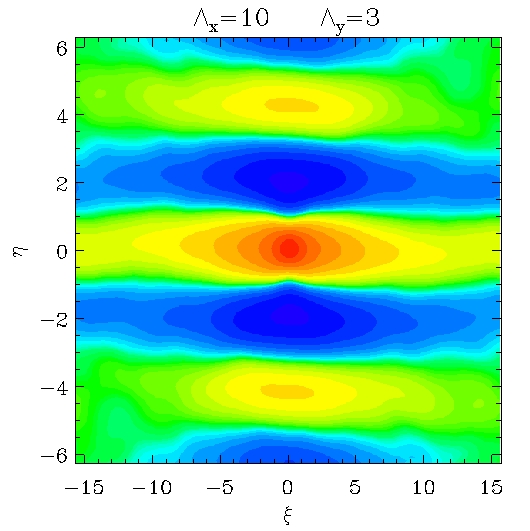}
%   \caption{}
   \label{fig:tmp_uu_66}
 \end{subfigure}
 \begin{subfigure}{0.32\textwidth}
\caption{}   \includegraphics[width=\textwidth]{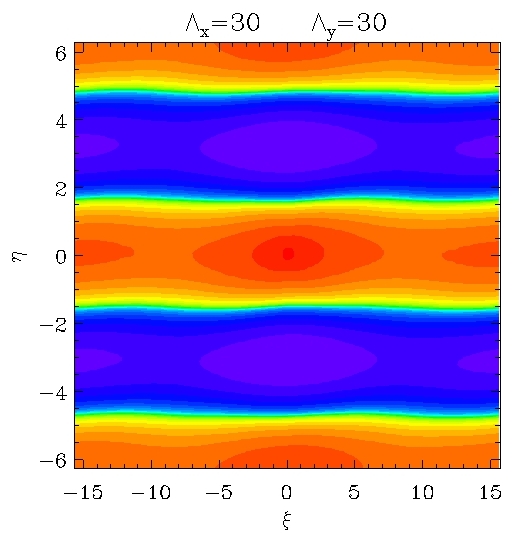}
%   \caption{}
   \label{fig:tmp_uu_3030}
 \end{subfigure}
\vskip -0.15truein 
\begin{subfigure}{0.98\textwidth}
\hskip 0.1truein
\includegraphics[width=\textwidth]{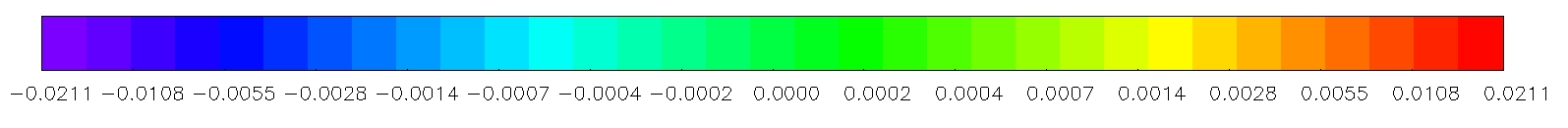}
%   \caption{}
   \label{fig:uu_cum_colbar}
 \end{subfigure}
  \caption{(Colour online): Second Cumulants \smtr{$c_{uu}(\xi,\eta,0.1,0.1)$} as defined in equation \ref{sec_cum_def} as a function of $\xi$ (horizontal) and $\eta$ (vertical).  
  Case A is shown (Table \ref{tbl:param}).}
  \label{fig:sec_cum_weak}
 \end{figure}

Figures~\ref{fig:spec_weak}c,d demonstrate that if the truncation is quasilinear in one direction then the cascade is severely inhibited in that direction, though transfer can occur in the orthogonal direction, because we have chosen to keep all the modes in this orthogonal direction as ``low" modes. Thus the energy transfer is artificially anisotropic on utilising such models. We note again here that ``minimal models" \citep{tfig2015} have been proposed for describing the evolution of shear instabilities that keep all spanwise modes and only the mean streamwise mode (equivalent to $\Lambda_x=0$, $\Lambda_y=$ all modes for GQL). What we have shown here is that, at least for this problem, this model performs poorly. \smt{One would expect the truncation that is quasilinear in the spanwise direction to perform poorly, as indeed it does.}

However, provided $\Lambda_x$, $\Lambda_y$ are both $\ne 0$ (shown in Figure~\ref{fig:spec_weak}e-h) then energy can be transported via non-local interactions to modes at small spatial scales, with the consequent improvement of the spectra in comparison with the Direct Numerical Simulation. This energy transport involves the successive scattering of high mode energy off the low modes into high modes with a different wavenumber. For example, if $\Lambda_x = \Lambda_y =3$ then a high mode with wavenumber $(6,7)$ can interact with a low mode with wavenumber $(1,1)$ to put energy into a high mode at wavenumbers $(7,8)$. In this non-local (in wavenumber space) interaction energy can be transferred to high wavenumbers to which the evolving shear flow is not directly unstable. Indeed successive interactions can redistribute energy among all the high modes as shown by the spectra. This is, we believe, {\it the key} advantage of the GQL approximation over the QL approximation.

\jbm{Figure~\ref{fig:sec_cum_weak} shows how well the various approximations perform in reproducing a second cumulant (or two-point correlation functions) of the RPC system for Case A. The second cumulant for the streamwise velocity $u$ is defined in terms of a spatial average over the two directions of translational invariance ($x$ and $y$), 
\begin{equation}
c_{uu}(\xi, \eta, z_1, z_2) = \frac{1}{L_x L_y}\int_{0}^{L_x} \int_0^{L_y} u^\prime(x, y, z_1)~ u^\prime(x+\xi, y+\eta, z_2) \,dx \,dy\,
\label{sec_cum_def}
\end{equation}
\smtr{where $u^\prime=u-U(z)$}.
In Figure~\ref{fig:sec_cum_weak} we plot, as a function of $\xi$ and $\eta$, a two dimensional slice at $z_1 = z_2 = 0.1$ of this four dimensional object.   Information contained in this piece of the second cumulant contributes to the  power spectra of Figure 8 --- however  the (logarithmically plotted) spectral power emphasizes the transfer of energy to small scales, while the second cumulant makes the ordering wavenumber readily apparent.  We note that the full second cumulant, being four dimensional, also contains information that is not present in the full three dimensional spectral power.}    
Recall from the discussion of Figure ~\ref{fig:dns}a that the major contribution to the streamwise velocity comes from the secondary flow, which, after a long transient with strong $(0,3)$ and $(2,3)$ components, settles into a dynamic $(0,2)$ and $(2,2)$ component. This pattern is selected in the second cumulant from DNS shown in figure~\ref{fig:sec_cum_weak}a . Interestingly the second cumulant from the QL approximation is dominated by the $(0,3)$ mode, as are some of the cumulants using GQL (see e.g. Figure~\ref{fig:sec_cum_weak}f for $\Lambda_x=3$, $\Lambda_y=10$). \smt{As stated earlier, this mode arises through the linear instability of the saturated shear flow, which is captured by QL. QL is however not able to capture the subsequent nonlinear transition to the wavy roll state.} The GQL integrations that (at least for this particular parameter choice) perform best in reproducing this second cumulant are shown in Figure~\ref{fig:sec_cum_weak}d,e,h --- interestingly those that include the most modes in the streamwise direction. \smt{This behaviour is non-monotonic for small truncation levels (the transition between the straight and wavy rolls is subtle and not always captured by severe truncations). However at moderate cutoffs GQL performs very well ---  see for example Figure~\ref{fig:tmp_uu_3030}.}

\begin{figure}
\begin{subfigure}{0.50\textwidth}
\caption{}
%   \includegraphics[width=\textwidth]{./new_f9a.jpg}
%\end{subfigure}
 %\begin{subfigure}{0.5\textwidth}
%\caption{}
   \includegraphics[width=\textwidth]{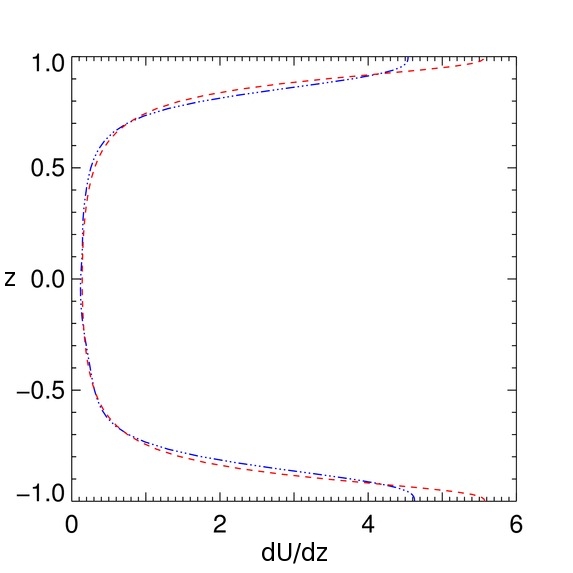}
 \end{subfigure}
\begin{subfigure}{0.5\textwidth}
\caption{}
   \includegraphics[width=\textwidth]{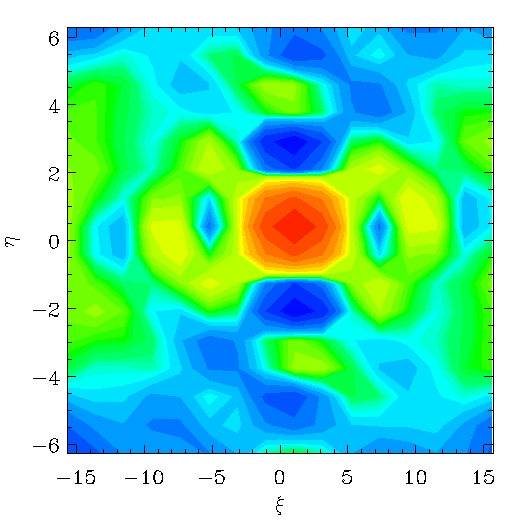}
\end{subfigure}
\caption{(Colour online) (a) Mean shears $dU/dz$ as a function of $z$ (vertical axis) for $Re=1300$, $Ro=-100$ (red), and for the under resolved case. \smtr{Note how the under-resolved case underestimates the wall shear stress and hence the enstrophy.} (b) Second cumulant for under-resolved case \smtr{which should be compared with the fully resolved case in Figure 9d and GQL with small cut-offs in Figures~9e and h}}
\label{fig:under}
 \end{figure}

\smt{We conclude this section by stressing that the GQL approximations performs significantly better than a poorly resolved Direct Numerical Simulation. For most values of the truncation utilised here, a DNS of the same resolution performed pseudo-spectrally will blow up owing to lack of resolution. We note that a $16 \times 16\times 64$ DNS is numerically stable for case A, but massively under-resolves the dissipation (yielding an enstrophy of $9.0$). Figure~\ref{fig:under} shows the shear and the second cumulant from the under-resolved run. Clearly at this resolution the scheme is incapable of reproducing both the mean shear near the boundaries and the spatial structure of the two-point correlation function. Thus, as expected, it performs significantly worse than a {\it more} severely truncated GQL run. }

\section{\label{sec:end}Discussion \& conclusions}

In this paper we have demonstrated that the Generalized Quasilinear approximation introduced by \citet{mct16} performs significantly better than the Quasilinear approximation for the problem of turbulent Rotating Couette flow at moderate Reynolds number. We have assessed the efficacy of the approximation in reproducing mean flows, spectra, dissipation rates and two-point correlation functions.  The approximation performs better than proposed minimal models that are quasilinear in the streamwise direction, even if only a few modes are kept as `low modes' in the streamwise and spanwise directions.  \jbm{That GQL shows a substantial and systematic improvement over QL for a three dimensional system with a forward cascade of energy from large to small scales is interesting, because nonlinear interactions that transfer energy only between high modes are dropped in the GQL approximation.  The result lays the foundation for Direct Statistical Simulation of three dimensional wall-bounded flows because the high modes can be closed statistically at second order, yielding a new class of subgrid models that can be systematically improved with increasing cutoff $\Lambda$.  Future work includes assessing the applicability of GQL for other wall-bounded systems such as Taylor-Couette and pipe flow, and completion of the program of Direct Statistical Simulation by implementation of the realizable GCE2 closure that is exact for GQL.}

\section*{Acknowledgements}
The authors would like to thank Greg Chini and Jeff Oishi. The computations were performed on ARC1 and ARC2, part of the High Performance Computing facilities at the University of Leeds, UK.  This work was supported in part by the US NSF under grant No. DMR-1306806 (JBM).
%\end{acknowledgements}

%
\bibliographystyle{jfm}
\bibliography{references}
\end{document}